\input{style/arxiv-general.cfg}
\documentclass[aoas,MSNbibl,nameyear,seceqn,secthm,dvips]{arximspdf}
\usepackage{url,breakurl}
\makeatletter
   \@ifpackageloaded{graphicx}{}{\usepackage{graphicx}}
\makeatother

\doi{10.1214/14-AOAS788} 
\volume{9}
\issue{1}
\pubyear{2015}
\firstpage{247}
\lastpage{274}
\docsubty{FLA}

\makeatletter
\newcommand{\rrvert}{\vert}
\newcommand{\llvert}{\vert}
\makeatother

\begin{document}
\begin{frontmatter}

\title{Inferring causal impact using Bayesian structural time-series models}
\runtitle{Bayesian causal impact analysis}

\begin{aug}
\author[A]{\fnms{Kay H.}~\snm{Brodersen}\corref{}\ead[label=e1]{kbrodersen@google.com}},
\author[A]{\fnms{Fabian}~\snm{Gallusser}\ead[label=e2]{gallusser@google.com}},
\author[A]{\fnms{Jim}~\snm{Koehler}\ead[label=e3]{jkoehler@google.com}},
\author[A]{\fnms{Nicolas}~\snm{Remy}\ead[label=e4]{nicolasremy@google.com}}
\and
\author[A]{\fnms{Steven L.}~\snm{Scott}\ead[label=e5]{stevescott@google.com}}
\runauthor{K. H. Brodersen et al.}
\affiliation{Google, Inc.}
\address[A]{Google, Inc.\\
1600 Amphitheatre Parkway\\
Mountain View, California 94043\\
USA\\
\printead{e1}\\
\phantom{E-mail:\ }\printead*{e2}\\
\phantom{E-mail:\ }\printead*{e3}\\
\phantom{E-mail:\ }\printead*{e4}\\
\phantom{E-mail:\ }\printead*{e5}}
\end{aug}

\received{\smonth{11} \syear{2013}}
\revised{\smonth{9} \syear{2014}}

%
\begin{abstract}
An important problem in econometrics and marketing is to infer the
causal impact that a designed market intervention has exerted on an
outcome metric over time.
This paper proposes to infer causal impact on the basis of a
diffusion-regression state-space model that predicts the
counterfactual market response in a synthetic control that would have
occurred had no
intervention taken place. In contrast to classical
difference-in-differences schemes, state-space models make it
possible to (i) infer the temporal evolution of attributable impact,
(ii) incorporate empirical priors on the parameters in a fully
Bayesian treatment, and (iii) flexibly accommodate multiple sources
of variation, including local trends, seasonality and the time-varying
influence of
contemporaneous covariates. Using a
Markov chain Monte Carlo algorithm for posterior inference, we
illustrate the statistical properties of our approach on simulated
data. We then demonstrate its practical utility by estimating the
causal effect of an online advertising campaign on search-related site
visits. We discuss the strengths and limitations of state-space models in
enabling causal attribution in those settings where a randomised
experiment is unavailable.
The CausalImpact R package provides an implementation of our approach.
\end{abstract}

%
\begin{keyword}
\kwd{Causal inference}
\kwd{counterfactual}
\kwd{synthetic control}
\kwd{observational}
\kwd{difference in differences}
\kwd{econometrics}
\kwd{advertising}
\kwd{market research}
\end{keyword}
\end{frontmatter}

\section{Introduction}
\label{sec:intro}

This article proposes an approach to inferring the causal impact of a
market intervention, such as a new product launch or the onset of an
advertising campaign. Our method generalises the widely used
difference-in-differences approach to the time-series setting by
explicitly modelling the counterfactual of a time series observed both
before and after the intervention. It improves on existing methods in
two respects: it provides a fully Bayesian time-series estimate for the effect;
and it uses model averaging to construct the most appropriate
synthetic control for modelling the counterfactual.
The \texttt{CausalImpact}  R package provides an implementation
of our approach (\url{http://google.github.io/CausalImpact/}).

Inferring the impact of market interventions is an important and
timely problem. Partly because of recent interest in big data,
many firms have begun to understand that a competitive advantage can
be had by systematically using impact measures to inform strategic
decision making. An example is the use of ``A$/$B experiments'' to
identify the most effective market treatments for the purpose of
allocating resources [\citet{danaherdetermining1996,seggiemeasurement2007,leeflangcreating2009,stewartmarketing2009}].

Here, we focus on measuring the impact of a discrete marketing event,
such as the release of a new product, the introduction of a new
feature, or the beginning or end of an advertising campaign, with the
aim of measuring the event's impact on a response metric of interest
(e.g., sales). The causal impact of a treatment is the difference
between the observed value of the response and the (unobserved) value
that would have been obtained under the alternative treatment, that is, the
effect of treatment on the treated
[\citet{antonakismaking2010,claveaurussowilliamson2012,coxcausal2001,heckmaneconometric2007,hitchcockall2004,%
hoovereconomic2012,kleinbergreview2011,morgancounterfactuals2007}, \citeauthor{rubinestimating1974}
(\citeyear{rubinestimating1974,rubinstatistical2007})].
In the
present setting the response variable is a time series, so the causal
effect of interest is the difference between the observed series and
the series that would have been observed had the intervention not
taken place.

A powerful approach to constructing the counterfactual is based on the
idea of combining a set of candidate predictor variables into a single
``synthetic control'' [\citet{abadieeconomic2003,abadiesynthetic2010}].
%
Broadly speaking, there are three sources of information available for
constructing an adequate synthetic control. The first is the
time-series behaviour of the response itself, prior to the
intervention. The second is the behaviour of other time series that
were predictive of the target series prior to the intervention.
Such control series can be based, for example, on the same product in a
different
region that did not receive the intervention or on a metric that
reflects activity in the industry as a whole. In
practice, there are often many such series available, and the
challenge is to pick the relevant subset to use as contemporaneous
controls.
%
This selection is done on the \emph{pre-treatment} portion of potential
controls; but their value for predicting the counterfactual lies in
their \emph{post-treatment} behaviour.
As long as the control series received no intervention
themselves, it is often reasonable to assume the relationship
between the treatment and the control series that existed prior to the
intervention to continue afterwards. Thus, a plausible estimate of
the counterfactual time series can be computed up to the point in time
where the relationship between treatment and controls can no longer be
assumed to be stationary, for example, because one of the controls received
treatment itself. In a Bayesian framework, a third source of
information for inferring the counterfactual is the available prior
knowledge about the model parameters, as elicited, for example, by
previous studies.
\begin{figure}[t]\vspace*{-9pt}

\includegraphics{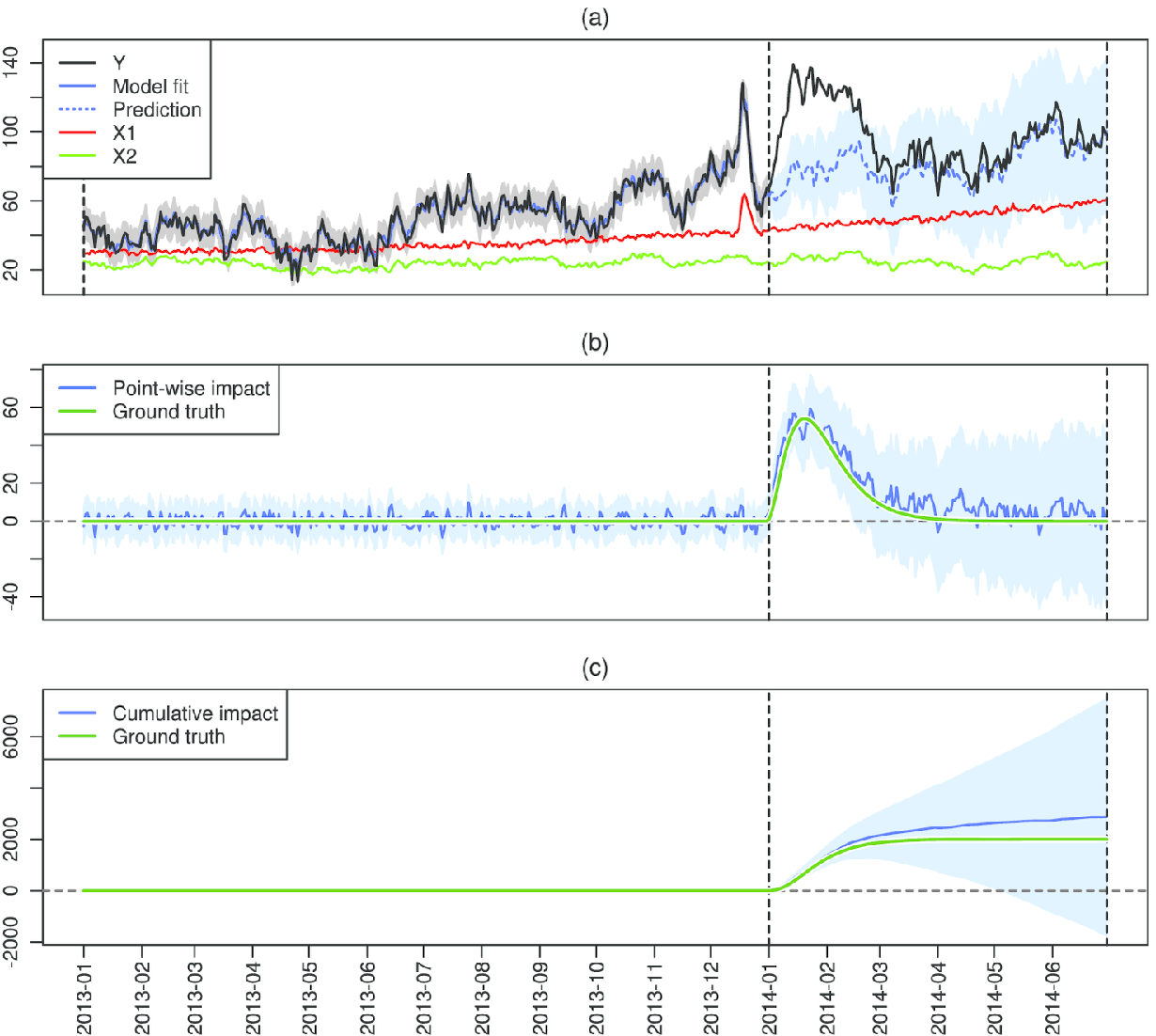}

\vspace*{-3pt}
\caption{Inferring causal impact through counterfactual predictions.
\textup{(a)}~Simulated trajectory of a treated market ($Y$) with an
intervention beginning in January 2014. Two other markets ($X_1$,
$X_2$) were not subject to the intervention and
allow us to construct a synthetic control [cf. \citet{abadieeconomic2003,abadiesynthetic2010}].
Inverting the state-space model
described in the main text yields a prediction of what would have
happened in $Y$ had the intervention not taken place (posterior
predictive expectation of the counterfactual with pointwise 95\%
posterior probability intervals). \textup{(b)}~The difference between
observed data and counterfactual predictions is the inferred
causal impact of the intervention. Here, predictions accurately
reflect the
true (Gamma-shaped) impact. A key characteristic of the
inferred impact series is the progressive widening of the
posterior intervals (shaded area). This effect emerges naturally
from the model structure and agrees with the intuition that
predictions should become increasingly uncertain as we look
further and further into the (retrospective) future. \textup{(c)}~Another
way of visualizing posterior inferences is by means of a
cumulative impact plot. It shows, for each day, the summed effect
up to that day. Here, the 95\% credible interval of the
cumulative impact crosses the zero-line about five months after
the intervention, at which point we would no longer declare
a significant overall effect.}\vspace*{-6pt}
\label{fig:intro:illustration}
\end{figure}

We combine the three preceding sources of information using a
state-space time-series model, where one component of state is a
linear regression on the contemporaneous predictors. The framework of
our model allows us to choose from among a large set of potential
controls by placing a spike-and-slab prior on the set of regression
coefficients and by allowing the model to average over the set of
controls [\citet{georgeapproaches1997}]. We then compute the
posterior distribution of the counterfactual time series given the
value of the target series in the pre-intervention period, along with
the values of the controls in the post-intervention period.
Subtracting the predicted from the observed response during the
post-intervention period gives a semiparametric Bayesian posterior
distribution for the causal effect
(Figure~\ref{fig:intro:illustration}).

\subsection*{Related work}

As with other domains, causal inference in marketing requires
subtlety. Marketing data are often observational and rarely follow
the ideal of a randomised design. They typically exhibit a low
signal-to-noise ratio. They are subject to multiple seasonal
variations, and they are often confounded by the effects of unobserved
variables and their interactions [for recent examples, see \citet{seggiemeasurement2007,stewartmarketing2009,leeflangcreating2009,%
takadamultiple1998,chanevaluating2010,lewisdoes2011,lewishere2011}, \citeauthor{vavermeasuring2011} (\citeyear{vavermeasuring2011,vaverperiodic2012})].

Rigorous causal inferences can be obtained through randomised
experiments, which are often implemented in the form of geo experiments\break
[\citeauthor{vavermeasuring2011} (\citeyear{vavermeasuring2011,vaverperiodic2012})]. Many
market interventions, however, fail to satisfy the requirements
of such approaches. For instance, advertising campaigns are
frequently launched across multiple channels, online and offline,
which precludes measurement of individual exposure. Campaigns are
often targeted at an entire country, and one country only, which
prohibits the use of geographic controls within that country.
Likewise, a campaign might be launched in several countries but at
different points in time. Thus, while a large control group may be
available, the treatment group often consists of no more than one
region or a few regions with considerable heterogeneity among them.

A standard approach to causal inference in such settings is based on a
linear model of the observed outcomes in the treatment and control
group before and after the intervention.
One can then
estimate the difference between (i)~the pre-post difference in the
treatment group and (ii)~the pre-post difference in the control group.
The assumption underlying such \emph{difference-in-differences} (DD)
designs is that the level of the control group provides an
adequate proxy for the level that would have been observed in the
treatment group in the absence of treatment
[see \citet{lestershortcomings1946,campbellexperimental1963,ashenfelterusing1985,cardminimum1993,angristempirical1999,atheyidentification2002,abadiesemiparametric2005,meyernatural1995,shadishexperimental2002,donaldinference2007,angristmostly2008,robinsonobserving2009,antonakismaking2010}].

DD designs have been limited in three ways. First, DD is traditionally
based on a static regression model that assumes i.i.d. data despite
the fact
that the design has a temporal component. When fit to serially
correlated data, static models yield overoptimistic inferences with
too narrow uncertainty intervals [see also \citet{bertrandhow2002},
\citeauthor{hansenasymptotic2007} (\citeyear{hansenasymptotic2007,hansengeneralized2007}),
\citet{solonestimating1984}].
Second, most DD analyses only consider two time
points: before and after the intervention. In practice, the manner in
which an effect evolves over time, especially its onset and decay
structure, is often a key question.

Third, when DD analyses \emph{are} based on time series, previous
studies have imposed restrictions on the way in which a synthetic
control is constructed from a set of predictor variables, which is
something we wish to avoid. For example, one strategy [\citet{abadieeconomic2003,abadiesynthetic2010}] has been to
choose a convex combination $(w_1, \ldots, w_J), w_j \geq0, \sum{w_j}
= 1$ of $J$ predictor time series in such a way that a vector of
pre-treatment variables (not time series) $X_1$ characterising the
treated unit before the intervention is matched most closely by the
combination of pre-treatment variables $X_0$ of the control units
w.r.t. a vector of importance weights $(v_1, \ldots, v_J)$. These
weights are themselves determined in such a way that the combination of
pre-treatment outcome time series of the control units most closely
matches the pre-treatment outcome time series of the treated unit. Such
a scheme relies on the availability of interpretable characteristics
(e.g., growth predictors), and it precludes nonconvex combinations of
controls when constructing the weight vector $W$. We prefer to select a
combination of control series without reference to external
characteristics and purely in terms of how well they explain the
pre-treatment outcome time series of the treated unit (while
automatically balancing goodness of fit and model complexity through
the use of regularizing priors).
Another idea [\citet{belloniprogram2013}] has been to use classical
variable-selection methods (such as the Lasso) to find a sparse set of
predictors. This approach, however, ignores posterior uncertainty about
both which predictors to use and their coefficients.

The limitations of DD schemes can be addressed by using state-space
models, coupled with highly flexible regression components, to explain
the temporal evolution of an observed outcome. State-space models distinguish
between a state equation that describes the transition of a set of
latent variables from one time point to the next and an observation
equation that specifies how a given system state translates into
measurements. This distinction makes them extremely flexible and powerful
[see \citet{leeflangcreating2009} for a discussion in the context of marketing
research].

The approach described in this paper inherits three main characteristics
from the state-space paradigm.
First, it allows us to flexibly accommodate different kinds
of assumptions about the latent state and emission processes underlying
the observed data, including local trends and seasonality.
Second, we use a fully Bayesian approach to inferring the temporal
evolution of
counterfactual activity and incremental impact. One advantage of this is
the flexibility with which posterior inferences can be summarised.
Third, we use a regression component that precludes a rigid commitment
to a particular set of controls by integrating out our posterior
uncertainty about the influence of each predictor as well as our
uncertainty about which predictors to include in the first place, which
avoids overfitting.

The remainder of this paper is organised as follows.
Section~\ref{sec:theory} describes the proposed model, its design
variations, the choice of diffuse empirical priors on hyperparameters,
and a stochastic algorithm for posterior inference based on Markov
chain Monte Carlo (MCMC). Section~\ref{sec:apps:synthetic} demonstrates
important features of the model using simulated data, followed by an
application in Section~\ref{sec:apps:empirical} to an advertising campaign
run by one of Google's advertisers. Section~\ref{sec:disc} puts our
approach into
context and discusses its scope of application.

\section{Bayesian structural time-series models}
\label{sec:theory}

Structural time-series models are state-space models for time-series
data. They can be defined in terms of a pair of equations
%
\begin{eqnarray}
\label{eq:observation} y_t &=& Z_t^{\mathrm{T}} {
\alpha}_t + \varepsilon_t,
\\
\label{eq:transition} {\alpha}_{t+1} &=& T_t {\alpha}_t
+ R_t \eta_t,
\end{eqnarray}
where $\varepsilon_t\sim\mathcal{N}(0, \sigma^2_t)$ and $\eta_t\sim
\mathcal{N}(0, Q_t)$ are independent of all other unknowns.
%
Equation~(\ref{eq:observation}) is the \textit{observation
equation}; it links the observed data $y_t$ to a latent $d$-dimensional
state vector ${\alpha}_t$.
Equation~(\ref{eq:transition}) is the
\textit{state equation}; it governs the evolution of
the state vector ${\alpha}_t$ through time.
In the present paper, $y_t$ is a scalar observation,
$Z_t$ is a $d$-dimensional output vector,
$T_t$ is a $d \times d$ transition matrix,
$R_t$ is a $d \times q$ control matrix,
$\varepsilon_t$ is a scalar observation error with noise variance $\sigma
_t$, and
$\eta_t$ is a $q$-dimensional system error with a $q \times q$
state-diffusion matrix $Q_t$,
where $q \le d$. Writing the error
structure of equation~(\ref{eq:transition}) as $R_t \eta_t$ allows us
to incorporate state components of less than full rank; a model for
seasonality will be the most important example.

Structural time-series models are useful in practice because they are
flexible and modular. They are flexible in the sense that a very
large class of models, including all ARIMA models, can be written in
the state-space form given by (\ref{eq:observation}) and (\ref{eq:transition}).
They are modular in the sense that the latent
state as well as the associated model matrices $Z_t, T_t, R_t$, and
$Q_t$ can be assembled from a library of component sub-models to
capture important features of the data. There are several widely
used state-component models for capturing the trend, seasonality
or effects of holidays.

A common approach is to assume the errors of different state-component
models to be independent (i.e., $Q_t$ is block-diagonal). The
vector ${\alpha}_t$ can then be formed by concatenating the individual state
components, while $T_t$ and $R_t$ become block-diagonal matrices.

The most important state component for the applications considered in
this paper
is a regression component that allows us to obtain counterfactual
predictions by constructing a synthetic control based on a combination
of markets that were not treated.
Observed responses from such markets
are important because they allow us to
explain variance components in the treated market that are not readily
captured by more generic seasonal sub-models.

This approach assumes that covariates are
unaffected by the effects of treatment. For example, an advertising
campaign run in the United States might spill over to Canada or the
United Kingdom. When assuming the absence of spill-over effects,
the use of such indirectly affected markets as controls would lead to
pessimistic inferences, that is, the effect of the campaign would be
underestimated
[cf. \citet{meyernatural1995}].\vadjust{\goodbreak}

\subsection{Components of state}
\label{sec:components-state}

\subsubsection*{Local linear trend}

The first component of our model is
a local linear trend, defined by the
pair of equations
%
\begin{eqnarray}
\mu_{t+1} &=& \mu_{t} + \delta_t +
\eta_{\mu, t},
\nonumber
\\[-8pt]
\label{eq:local-linear-trend}
\\[-8pt]
\nonumber
\delta_{t+1} &= &\delta_{t} + \eta_{\delta, t},
\end{eqnarray}
where $\eta_{\mu, t} \sim\mathcal{N}(0, \sigma^2_\mu)$ and
$\eta_{\delta, t} \sim\mathcal{N}(0, \sigma^2_\delta)$. The $\mu_t$
component is the value of the trend at time $t$. The $\delta_t$
component is the expected increase in $\mu$ between times $t$ and
$t+1$, so it can be thought of as the \emph{slope} at time $t$.

The local linear trend model is a popular choice for modelling trends
because it quickly adapts to local variation, which is desirable when
making short-term predictions. This degree of flexibility may not be
desired when making longer-term predictions, as such predictions often
come with implausibly wide uncertainty intervals.

There is a generalisation of the local linear trend model where the
slope exhibits stationarity instead of obeying a random walk. This model
can be written as
%
\begin{eqnarray}
\mu_{t+1} &=& \mu_t + \delta_t +
\eta_{\mu, t},
\nonumber
\\[-8pt]
\label{eq:generalized-local-linear-trend}
\\[-8pt]
\nonumber
\delta_{t+1} &=& D + \rho(\delta_{t} - D) +
\eta_{\delta, t},
\end{eqnarray}
where the two components of $\eta$ are independent. In this model, the
slope of the time trend exhibits $\mathrm{AR}(1)$ variation around a long-term
slope of $D$. The parameter $|\rho| < 1$ represents the learning rate
at which the local trend is updated. Thus, the model balances
short-term information with information from the distant past.

\subsubsection*{Seasonality}

There are several commonly used state-component models to capture
seasonality. The most frequently used model in the time domain is
%
\begin{equation}
\label{eq:seasonal}
\gamma_{t+1} = -\sum_{s = 0}^{S-2}
\gamma_{t-s} + \eta_{\gamma, t},
\end{equation}
where $S$ represents the number of seasons and $\gamma_t$ denotes
their joint contribution to the observed response $y_t$. The state in
this model consists of the $S-1$ most recent seasonal effects, but the
error term is a scalar, so the evolution equation for this state model
is less than full rank. The mean of $\gamma_{t+1}$ is such that the
total seasonal effect is zero when summed over $S$ seasons. For
example, if we set $S=4$ to capture four seasons per year, the mean of
the \emph{winter} coefficient will be $-1 \times(\mathit{spring} +
\mathit{summer} + \mathit{autumn})$. The part of the transition matrix
$T_t$ representing
the seasonal model is an $S - 1 \times S - 1$ matrix with $-1$'s
along the top row, 1's along the subdiagonal and 0's elsewhere.

The preceding seasonal model can be generalised to allow for multiple
seasonal components with different periods. When modelling daily data,
for example, we might wish to allow for an $S=7$ day-of-week effect, as
well as an $S=52$ weekly annual cycle. The latter can be handled by
setting $T_t = I_{S-1}$, with zero variance on the error term, when $t$
is not the start of a new week, and setting $T_t$ to the usual seasonal
transition matrix, with nonzero error variance, when $t$ is the start
of a new
week.

\subsubsection*{Contemporaneous covariates with static coefficients}

Control time series that received no treatment are
critical to our method for obtaining accurate counterfactual
predictions since they account for variance components that are shared
by the series, including, in particular, the effects of other unobserved
causes otherwise unaccounted for by the model.
%
A natural way of including control series in the model is through
a linear regression. Its coefficients can be static
or time-varying.


A \emph{static} regression can be written in state-space form by
setting $Z_t = {\beta}^{\mathrm{T}} \mathbf{x}_t$ and ${\alpha}_t = 1$.
One advantage of working in a fully Bayesian treatment is that
we do not need to commit to a fixed set of covariates. The
spike-and-slab prior described in Section~\ref{sec:posterior-sampling}
allows us to integrate out our posterior uncertainty about which
covariates to include and how strongly they should influence our
predictions, which avoids overfitting.

All covariates are assumed to be contemporaneous; the present model does
not infer on a potential lag between treated and untreated time
series. A known lag, however, can be easily incorporated by shifting
the corresponding regressor in time.

\subsubsection*{Contemporaneous covariates with dynamic coefficients}

An alternative to the above is a regression component with
\emph{dynamic} regression coefficients to account for time-varying
relationships [e.g., \citet{banerjeemodeling2007,westharr1997}].
Given covariates $j = 1, \ldots, J$, this introduces the dynamic
regression component
%
\begin{eqnarray}
\mathbf{x}_t^{\mathrm{T}} {\beta}_t &= &\sum
_{j=1}^J{x_{j,t} \beta _{j,t}},
\nonumber
\\[-8pt]
\label{eq:dynamic-regression}
\\[-8pt]
\nonumber
\beta_{j,t+1} &=& \beta_{j,t} + \eta_{\beta, j, t},
\end{eqnarray}
where $\eta_{\beta, j, t}\sim\mathcal{N}(0,
\sigma^2_{\beta_j})$. Here, $\beta_{j,t}$ is
the coefficient for the $j$th control series and
$\sigma_{\beta_j}$ is the standard deviation of its associated random
walk. We can write the dynamic regression component in
state-space form by setting $Z_t = \mathbf{x}_t$ and ${\alpha}_t =
{\beta}_t$
and by setting\vspace*{1pt} the corresponding part of the transition matrix to $T_t
= I_{J \times J}$, with $Q_t = \operatorname{diag}
(\sigma_{\beta_j}^2)$.

\subsubsection*{Assembling the state-space model}

Structural time-series models allow us to examine the time series at
hand and flexibly
choose appropriate components for trend, seasonality, and either static or
dynamic regression for the controls. The presence
or absence of seasonality, for example, will usually be obvious by inspection.
%
A more subtle question is whether to choose static or dynamic
regression coefficients.

When the relationship between controls and treated unit has been stable
in the past, static coefficients are an attractive option. This is
because a spike-and-slab prior can be implemented efficiently within a
forward-filtering, backward-sampling framework. This makes it possible
to quickly identify a sparse set of covariates even from tens or
hundreds of potential variables [\citet{scotvari2013}]. Local
variability in the treated time series is captured by the dynamic local
level or dynamic linear trend component.
Covariate stability is typically high when the available covariates are
close in nature to the treated metric.
The empirical analyses presented in this paper, for example, will be
based on a static regression component (Section~\ref{sec:apps:empirical}).
This choice provides a reasonable compromise between capturing local
behaviour and accounting for regression effects.

An alternative would be to use dynamic regression coefficients, as we
do, for instance, in our analyses of simulated data (Section~\ref{sec:apps:synthetic}). Dynamic coefficients are useful when the linear
relationship between treated metrics and controls is believed to change
over time. There are a number of ways of reducing the computational
burden of dealing with a potentially large number of dynamic coefficients.
One option is to resort to dynamic latent factors, where one uses
$\mathbf{x}
_t = B{\mathbf{u}}_t + {\bolds{\nu}}_t$ with $\operatorname{dim}({\mathbf{u}}_t) \ll J$ and
uses ${\mathbf{u}}_t$ instead of $\mathbf{x}_t$ as part of $Z_t$ in (\ref{eq:observation}), coupled with an AR-type model for ${\mathbf{u}}_t$ itself.
Another option is latent thresholding regression, where one uses a
dynamic version of the spike-and-slab prior as in \citet{nakajimabayesian2013}.

\begin{figure}

\includegraphics{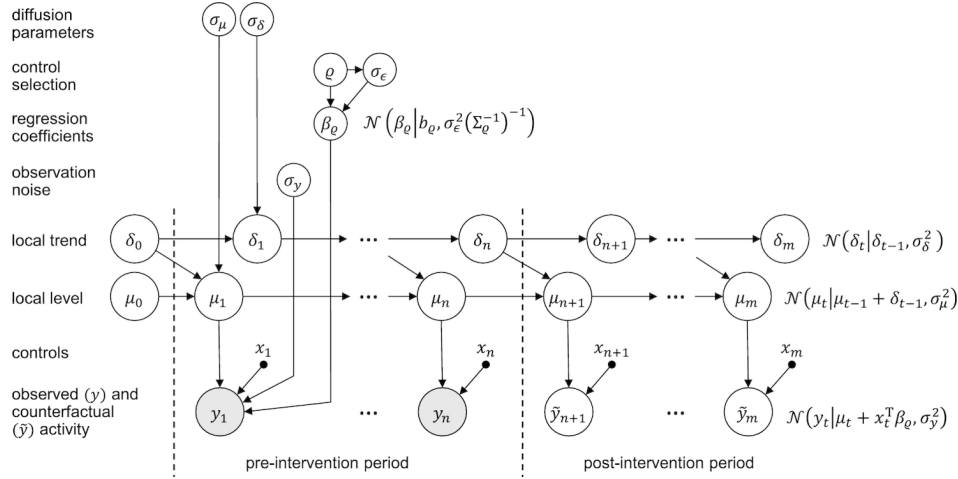}

\caption{Graphical model for the static-regression variant of the proposed
state-space model. Observed market activity $\mathbf{y}_{1\dvtx n} = (y_1,
\ldots,
y_n)$ is modelled as
the result of a latent state plus Gaussian observation noise with
error standard deviation $\sigma_y$. The state $\alpha_t$ includes a local
level $\mu_t$, a local linear trend $\delta_t$,  and a set of contemporaneous
covariates $\mathbf{x}_t$, scaled by regression coefficients ${\beta
}_{\varrho}$.
State
components are assumed to evolve according to independent Gaussian
random walks with fixed standard deviations $\sigma_\mu$ and
$\sigma_\delta$ (conditional-dependence arrows shown for the first
time point only). The model includes empirical priors on these
parameters and the initial states. In an alternative formulation,
the regression coefficients ${\beta}$ are themselves subject to
random-walk diffusion (see main text).
Of principal interest is the posterior predictive density
over the unobserved counterfactual responses $\tilde{y}_{n+1},
\ldots, \tilde{y}_m$. Subtracting these from the actual observed
data $y_{n+1}, \ldots, y_m$ yields a probability density over the
temporal evolution of causal impact.}
\label{fig:theory:model}
\end{figure}

The state-component models are assembled independently, with each
component providing an additive contribution to $y_t$.
Figure~\ref{fig:theory:model} illustrates this process assuming a
local linear trend paired with a static regression component.

\subsection{Prior distributions and prior elicitation}
\label{sec:posterior-sampling}

Let ${\theta}$ generically denote the set of all model parameters and
let ${\alpha}= ({\alpha}_1, \ldots, {\alpha}_m)$ denote the full state
sequence. We
adopt a Bayesian approach to inference by specifying a prior
distribution $p({\theta})$ on the model parameters as well as a
distribution $p({\alpha}_0 |{\theta})$ on the initial state values.
We may
then sample from $p({\alpha}, {\theta}|\mathbf{y})$ using MCMC.

Most of the models in Section~\ref{sec:components-state} depend solely
on a small set of variance parameters that govern the diffusion of the
individual state components. A typical prior distribution
for such a variance is
%
\begin{equation}
\label{eq:typical-variance}
\frac{1}{\sigma^2} \sim\mathcal{G} \biggl(\frac{\nu}{2},
\frac
{s}{2} \biggr),
\end{equation}
where $\mathcal{G} (a, b)$ is the Gamma distribution with
expectation $a/b$. The
prior parameters can be interpreted as a prior sum of squares $s$, so
that $s/\nu$ is a prior estimate of~$\sigma^2$, and $\nu$ is the
weight, in units of prior sample size, assigned to the prior estimate.

We often have a weak default prior belief that the incremental errors
in the state process are small, which we can formalise by choosing
small values of $\nu$ (e.g., 1) and small values of $s/\nu$. The
notion of ``small'' means different things in different models; for the
seasonal and local linear trend models our default priors are
$1/\sigma^2 \sim\mathcal{G}(10^{-2}, 10^{-2} s^2_y)$, where $s^2_y =
\sum_t (y_t - \bar y)^2 / (n-1)$ is the sample variance of the target
series. Scaling by the sample variance is a minor violation of the
Bayesian paradigm, but it is an effective means of choosing a
reasonable scale for the prior. It is similar to the popular
technique of scaling the data prior to analysis, but we prefer to do
the scaling in the prior so we can model the data on its original
scale.

When faced with many potential controls, we prefer
letting the model choose an appropriate set. This can be
achieved by placing a spike-and-slab prior over coefficients
[\citeauthor{georgevariable1993} (\citeyear{georgevariable1993,georgeapproaches1997}), \citet{polsondata2011,scotvari2013}]. A spike-and-slab prior combines point mass at zero
(the ``spike''), for an unknown subset of zero coefficients, with a
weakly informative distribution on the complementary set of nonzero
coefficients (the ``slab''). Contrary to what its name might suggest,
the ``slab'' is usually not completely flat, but rather a Gaussian with
a large variance. Let ${\varrho}= (\varrho_1, \ldots, \varrho_J)$, where
$\varrho_j = 1$ if $\beta_j \ne0$ and $\varrho_j = 0$ otherwise. Let\vspace*{-1.5pt}
${\beta}_\varrho$ denote the nonzero elements of the vector ${\beta
}$ and
let $\Sigma^{-1}_\varrho$ denote\vspace*{1pt} the rows and columns of $\Sigma^{-1}$
corresponding to nonzero entries in ${\varrho}$. We can then factorise the
spike-and-slab prior as
%
\begin{equation}
\label{eq:spike-slab-general}
p\bigl({\varrho}, {\beta}, 1/\sigma^2_\varepsilon
\bigr) = p({\varrho}) p\bigl(\sigma^2_\varepsilon| {\varrho}\bigr)
p\bigl({\beta}_{\varrho}| {\varrho}, \sigma^2_\varepsilon
\bigr).
\end{equation}
The spike portion of (\ref{eq:spike-slab-general}) can be an
arbitrary distribution over $\{0, 1\}^J$ in principle; the most common
choice in practice is a product of independent Bernoulli distributions,
%
\begin{equation}
\label{eq:spike}
p({\varrho}) = \prod_{j=1}^J
\pi_j^{\varrho_j}(1 - \pi_j)^{1 -
\varrho_j},
\end{equation}
where $\pi_j$ is the prior probability of regressor $j$ being included
in the model.

Values for $\pi_j$
can be elicited by asking about the \emph{expected model size} $M$, and
then setting all $\pi_j = M/J$. An alternative is to use a more
specific set
of values~$\pi_j$. In particular, one might
choose to set certain $\pi_j$ to either~1 or~0 to force the
corresponding variables into or out of the model.
Generally, framing the prior in terms of expected model size has the
advantage that the model can adapt to growing numbers of predictor
variables without having to switch to a hierarchical prior
[\citet{scottbayes2010}].

For the ``slab'' portion of the prior we use a conjugate normal-inverse
Gamma distribution,
%
\begin{eqnarray}\label{eq:slab-normal}
{\beta}_{\varrho}|\sigma^2_\varepsilon &\sim &  \mathcal{N}
\bigl({\mathbf b}_{\varrho}, \sigma^2_\varepsilon\bigl(\Sigma
^{-1}_{\varrho}\bigr)^{-1} \bigr),
\\
\label{eq:slab-variance}
\frac{1}{\sigma^2_\varepsilon} & \sim & \mathcal{G} \biggl(\frac{\nu
_\varepsilon}{2}, \frac {s_{\varepsilon}}{2}
\biggr).
\end{eqnarray}
The vector ${\mathbf b}$ in equation~(\ref{eq:slab-normal}) encodes our
prior expectation about the value of each element of ${\beta}$. In
practice, we
usually set ${\mathbf b} = 0$. The prior\vspace*{1pt} parameters in
equation~(\ref{eq:slab-variance}) can be elicited by asking about
the expected $R^2 \in[0, 1]$ as well as the number of
observations worth of weight $\nu_\varepsilon$ the prior estimate should
be given. Then $s_\varepsilon= \nu_\varepsilon(1 - R^2) s^2_y$.

The final prior parameter in (\ref{eq:slab-normal}) is $\Sigma^{-1}$,
which, up to a scaling factor, is the prior precision over ${\beta}$
in the full model, with all variables included. The total information
in the covariates is $X^{\mathrm{T}}X$, and so $\frac{1}{n}
X^{\mathrm{T}}X$ is the average
information in a single observation. Zellner's $g$-prior\vspace*{1pt}
[\citet{zell1986,chipgeormccuclydfoststin2001,liangetal2008}]
sets $\Sigma^{-1}= \frac{g}{n} X^{\mathrm{T}}X$, so that $g$ can be
interpreted as
$g$ observations worth of information. Zellner's prior becomes
improper when $X^{\mathrm{T}}X$ is not positive definite; we therefore ensure
propriety by averaging $X^{\mathrm{T}}X$ with its diagonal,
%
\begin{equation}
\label{eq:siginv}
\Sigma^{-1}= \frac{g}{n} \bigl\{ wX^{\mathrm{T}}X
+ (1-w) \operatorname{diag} \bigl(X^{\mathrm{T}}X \bigr) \bigr\}
\end{equation}
with default values of $g = 1$ and $w = 1/2$.
Overall, this prior specification provides a broadly useful default
while providing considerable flexibility in those cases where more
specific prior information is available.

\subsection{Inference}
\label{sec:mcmc}

Posterior inference in our model can be broken down into three pieces.
First, we simulate draws of the model parameters $\theta$ and the
state vector ${\alpha}$ given the observed data $\mathbf{y}_{1 \dvtx  n}$ in
the training
period. Second, we use the posterior simulations to simulate from the
posterior predictive distribution $p({\tilde\mathbf{y}}_{n+1\dvtx m}
|\mathbf{y}_{1\dvtx n})$
over the
counterfactual time series ${\tilde\mathbf{y}}_{n+1\dvtx m}$ given the observed
pre-intervention activity $\mathbf{y}_{1\dvtx n}$. Third, we use the posterior
predictive samples to compute the posterior distribution of the
pointwise impact $y_t - \tilde y_t$ for each $t = 1, \ldots, m$. We use
the same
samples to obtain the posterior distribution of cumulative impact.

\subsubsection*{Posterior simulation}

We use a Gibbs sampler to simulate a sequence $(\theta,
{\alpha})^{(1)},
(\theta, {\alpha})^{(2)}, \ldots$ from a Markov chain whose stationary
distribution is $p(\theta, {\alpha}|\mathbf{y}_{1\dvtx n})$. The sampler
alternates between
a \emph{data-augmentation} step that simulates from $p({\alpha
}|\mathbf{y}_{1\dvtx n},
\theta)$ and a \emph{parameter-simulation} step that simulates from
$p(\theta| \mathbf{y}_{1\dvtx n}, {\alpha})$.

The data-augmentation step uses the posterior simulation algorithm from
\citet{durbkoop2002}, providing an improvement over the earlier
forward-filtering,
backward-sampling algorithms by \citet{cartkohn1994},
\citet{fruh1994}, and \citet{dejoshep1995}. In brief, because
$p(\mathbf{y}_{1\dvtx n},
{\alpha}|\theta)$ is jointly multivariate normal, the variance of
$p({\alpha}|
\mathbf{y}_{1\dvtx n}, \theta)$ does not depend on $\mathbf{y}_{1\dvtx n}$. We
can therefore
simulate $(\mathbf{y}_{1\dvtx n}^*,
{\alpha}^*) \sim p(\mathbf{y}_{1\dvtx n}, {\alpha}|\theta)$ and subtract
$\textrm
{E}({\alpha}^*|\mathbf{y}_{1\dvtx n}^*, \theta)$
to obtain zero-mean noise with the correct variance. Adding $\textrm
{E}({\alpha}|
\mathbf{y}_{1\dvtx n}, \theta)$ restores the correct mean, which completes
the draw.
The required expectations can be computed using the Kalman filter and a
\emph{fast mean smoother} described in detail by \citet{durbkoop2002}.
The result is a direct simulation from $p({\alpha}|\mathbf{y}_{1\dvtx n},
\theta)$
in an
algorithm that is linear in the total (pre- and post-intervention)
number of time points ($m$) and quadratic
in the dimension of the state space ($d$).

Given the draw of the state, the parameter draw is straightforward for
all state components other than the static regression coefficients
$\beta$. All state components that exclusively depend on variance
parameters can translate their draws back to error terms $\eta_t$ and
accumulate sums of squares of $\eta$, and, because of conjugacy with
equation~(\ref{eq:typical-variance}), the posterior distribution will
remain Gamma distributed.

The draw of the static regression coefficients $\beta$ proceeds as
follows. For each $t = 1, \ldots, n$ in the pre-intervention period,
let $\dot{y}_t$ denote
$y_t$ with the contributions from the other state components
subtracted away, and let $\dot{\mathbf{y}}_{1\dvtx n} = (\dot{y}_1, \ldots,
\dot{y}_n)$. The
challenge is to simulate from $p(\varrho, \beta, \sigma^2_\varepsilon
|\dot{\mathbf{y}}_{1\dvtx n})$, which we can factor into $p(\varrho|\mathbf{y}_{1\dvtx n})
p(1/\sigma^2_\varepsilon|\varrho, \dot{\mathbf{y}}_{1\dvtx n}) p(\beta
|\varrho,
\sigma_\varepsilon, \dot{\mathbf{y}}_{1\dvtx n})$.\vspace*{1pt} Because of conjugacy, we
can integrate
out $\beta$ and $1/\sigma^2_\varepsilon$ and be left with
%
\begin{equation}
\label{eq:inclusion-indicator-posterior}
\varrho| \dot{\mathbf{y}}_{1\dvtx n} \sim C(\dot{\mathbf{y}}_{1\dvtx n})
\frac{|\Sigma^{-1}_\varrho|^{{1}/{2}}
}{
|V^{-1}_\varrho|^{{1}/{2}}
} \frac{
p(\varrho)
}{S_\varrho^{({N}/{2})-1}},
\end{equation}
where $C(\dot\mathbf{y}_{1\dvtx n})$ is an unknown normalizing constant.
The sufficient
statistics in equation~(\ref{eq:inclusion-indicator-posterior}) are
\begin{eqnarray*}
V_\varrho^{-1} & =& \bigl(X^{\mathrm{T}}X \bigr)_\varrho+
\Sigma ^{-1}_\varrho, \qquad  \tilde\beta_\varrho =
\bigl(V^{-1}_\varrho\bigr)^{-1}\bigl(\mathbf
{X}_\varrho^{\mathrm{T}}\dot\mathbf{y}_{1\dvtx n} +
\Sigma^{-1}_\varrho b_\varrho\bigr),
\\
N &=& \nu_\varepsilon+ n,\qquad  S_\varrho = s_\varepsilon + \dot
\mathbf{y}_{1\dvtx n}^{\mathrm{T}}\dot\mathbf{y}_{1\dvtx n} +
b_\varrho^{\mathrm{T}}\Sigma^{-1}_\varrho
b_\varrho - \tilde\beta_\varrho^{\mathrm{T}}V_\varrho^{-1}
\tilde\beta _\varrho.
\end{eqnarray*}
To sample from (\ref{eq:inclusion-indicator-posterior}), we use a
Gibbs sampler that draws each $\varrho_j$ given all other
$\varrho_{-j}$. Each full-conditional is easy to evaluate because
$\varrho_j$ can only assume two possible values. It should be noted
that the dimension of all matrices in
(\ref{eq:inclusion-indicator-posterior}) is $\sum_j \varrho_j$,
which is small if the model is truly sparse. There are many matrices
to manipulate, but because each is small, the overall\vspace*{1pt} algorithm is fast.
Once the draw of $\varrho$ is complete, we sample directly from
$p(\beta,
1/\sigma^2_\varepsilon| \varrho, \dot\mathbf{y}_{1\dvtx n})$ using standard
conjugate formulae. For an alternative that may be
even more computationally efficient, see \citet{ghosclyd2011}.

\subsubsection*{Posterior predictive simulation}

While the posterior over model parameters and states $p(\theta,
\bolds{\alpha} | \mathbf{y}_{1\dvtx n})$ can be of interest in its own
right, causal
impact analyses are primarily concerned with the posterior incremental
effect,
%
\begin{equation}\label{eqn:theory:postpred}
p (\tilde{\mathbf{y}}_{n+1\dvtx m} \vert \mathbf{y}_{1\dvtx n},
\mathbf{x}_{1\dvtx m}).
\end{equation}
%
%
As shown by its indices, the density in
equation~(\ref{eqn:theory:postpred}) is defined precisely for that
portion of the time series which is unobserved: the counterfactual
market response $\tilde{y}_{n+1}, \ldots, \tilde{y}_m$ that would
have been observed in the treated market, after the intervention,
in the absence of treatment.

It is also worth emphasising that the density is conditional on the
observed data (as well as the priors) and only on these, that is, on
activity in the treatment market before the beginning of the
intervention as well as activity in all control markets both before
and during the intervention. The density is \emph{not} conditioned on
parameter estimates or the inclusion or exclusion of covariates with
static regression coefficients, all of which have been integrated out.
Thus, through Bayesian model averaging, we
commit neither to any particular
set of covariates, which helps avoid an arbitrary selection, nor to
point estimates of their coefficients, which prevents
overfitting.

The posterior predictive density in (\ref{eqn:theory:postpred}) is
defined as a coherent (joint) distribution over all counterfactual
data points, rather than as a collection of pointwise univariate
distributions. This ensures that we correctly propagate the serial
structure determined on pre-intervention data to the trajectory of
counterfactuals. This is crucial, in particular, when forming summary
statistics, such as the cumulative effect of the intervention on the
treatment market.

Posterior inference was implemented in C${++}$ with an {R} interface. Given
a typically-sized data set with $m = 500$ time points, $J = 10$
covariates,  and $10{,}000$ iterations (see Section~\ref{sec:apps:empirical} for an example), this implementation takes less than 30~seconds to
complete on a standard computer, enabling near-interactive analyses.

\subsection{Evaluating impact}
\label{sec:evaluating-impact}

Samples from the posterior predictive distribution over counterfactual
activity can be readily used to obtain samples from the posterior
causal effect, that is, the quantity we are typically interested in.
For each draw $\tau$ and for each time point $t = n+1, \ldots, m$, we set
%
\begin{equation}\label{eqn:theory:causal_samples}
\phi_t^{(\tau)} := y_t - \tilde{y}_t^{(\tau)},
\end{equation}
yielding samples from the approximate
posterior predictive density of the effect attributed to the
intervention.

In addition to its pointwise impact, we often wish to understand the
cumulative effect of an intervention over time.
One of the main advantages of a sampling approach to
posterior inference is the flexibility and ease with which such
derived inferences can be obtained. Reusing the impact samples
obtained in (\ref{eqn:theory:causal_samples}), we compute for each
draw $\tau$
%
\begin{equation}\label{eqn:cumlift}
\sum_{t' = n+1}^t \phi_{t'}^{(\tau)} \qquad
\forall t = n+1, \ldots, m.
\end{equation}
The preceding \emph{cumulative sum} of causal increments is a useful quantity
when
$y$ represents a \emph{flow} quantity, measured over an interval of
time (e.g.,
a day), such
as the number of searches, sign-ups, sales, additional installs or new
users. It becomes uninterpretable when $y$ represents a \emph{stock}
quantity, usefully defined only for a point in time, such as the total
number of clients, users or subscribers.
In this case we might instead choose, for each $\tau$, to draw a
sample of
the posterior \emph{running average}
effect following the intervention,
%
\begin{equation}
\frac{1}{t-n} \sum_{t' = n+1}^t
\phi_{t'}^{(\tau)} \qquad \forall t = n+1, \ldots, m.
\end{equation}
Unlike the cumulative effect in (\ref{eqn:cumlift}), the running average
is always interpretable, regardless of whether it refers to a flow
or a stock. However, it is more context-dependent on the length of the
post-intervention period under consideration. In particular, under the
assumption of a true
impact that grows quickly at first and then declines to zero, the
cumulative impact approaches its true total value (in expectation) as we
increase the
counterfactual forecasting period, whereas the average impact will eventually
approach zero (while, in contrast, the probability intervals diverge
in both cases, leading to more and more uncertain inferences as the
forecasting period increases).

\section{Application to simulated data}
\label{sec:apps:synthetic}

\begin{figure}

\includegraphics{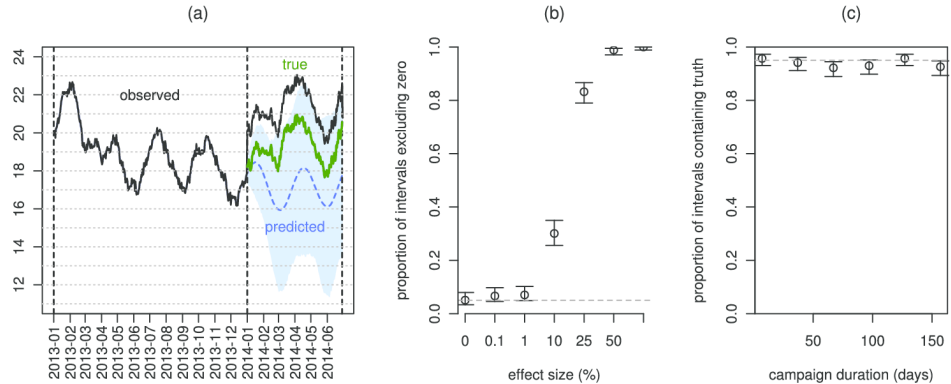}

\caption{Adequacy of posterior uncertainty. \textup{(a)}~Example of one of
the 256~data sets created to assess estimation accuracy. Simulated
observations (black) are based on two contemporaneous covariates,
scaled by time-varying coefficients plus a time-varying local
level (not shown). During the campaign period, where the data are
lifted by an effect size of 10\%, the plot shows the posterior
expectation of counterfactual activity (blue), along with its pointwise
central 95\% credible intervals (blue shaded area), and, for
comparison, the true counterfactual (green).
\textup{(b)}~Power curve. Following repeated application of the model to
simulated data, the plot shows the empirical frequency of
concluding that a causal effect was present, as a function of true
effect size, given a post-intervention period of 6~months. The curve
represents sensitivity in those parts of
the graph where the true effect size is positive, and 1---specificity where the true effect size is zero. Error bars
represent 95\% credible intervals for the true sensitivity, using
a uniform $\operatorname{Beta}(1, 1)$ prior.
\textup{(c)}~Interval coverage. Using an effect size of 10\%, the plot
shows the proportion of simulations in which the pointwise
central 95\% credible interval contained the true impact, as a
function of campaign duration. Intervals should contain ground
truth in 95\% of simulations, however much uncertainty its
predictions may be associated with. Error bars represent 95\%
credible intervals.}
\label{fig:apps:synthetic:inference}
\end{figure}

To study the characteristics of our approach, we analysed simulated
(i.e., computer-generated) data
across a series of independent simulations.
%
Generated time series started on 1~January 2013 and ended on
30~June 2014, with a perturbation beginning on 1~January 2014. The
data were simulated using a dynamic regression component with two covariates
whose coefficients evolved according to independent random walks,
$\beta_t \sim\mathcal{N}(\beta_{t-1}, 0.01^2)$, initialised at
$\beta_0 = 1$. The covariates themselves were simple sinusoids with
wavelengths of 90~days and 360~days, respectively. The latent state
underlying the observed data was generated using a local level that
evolved according to a random walk, $\mu_t \sim\mathcal{N}(\mu_{t-1},
0.1^2)$, initialised at $\mu_0 = 0$. Independent observation noise was
sampled using $\varepsilon_t \sim\mathcal{N}(0, 0.1^2)$. In summary,
observations $y_t$ were generated using
\[
y_t = \beta_{t,1} z_{t,1} + \beta_{t,2}
z_{t,2} + \mu_t + \varepsilon _{t}.
\]
To simulate the effect of advertising, the post-intervention portion
of the preceding series was multiplied by $1+e$, where $e$ (not to be
confused with $\varepsilon$) represented the true effect size specifying
the (uniform) relative lift during the campaign period. An example is
shown in Figure~\ref{fig:apps:synthetic:inference}(a).

%

\subsection*{Sensitivity and specificity}

To study the properties of our model, we began by considering under
what circumstances we successfully detected a causal effect,
that is, the statistical power or sensitivity of our approach. A related
property is the probability of \emph{not} detecting an
absent impact, that is, specificity. We repeatedly generated
data, as described above, under different true effect sizes. We then
computed the posterior
predictive distribution over the counterfactuals and recorded whether or
not we would have concluded a causal effect.

For each of the effect sizes 0\%, 0.1\%, 1\%, 10\% and 100\%, a total
of $2^8 = 256$ simulations were run.
%
This number was chosen simply on the grounds that it provided
reasonably tight intervals around the reported summary statistics
without requiring excessive amounts of computation.
In each simulation, we concluded that a
causal effect was present if and only if the central 95\% posterior
probability interval of the cumulative effect excluded zero.


The model used throughout this section comprised two structural blocks.
The first one was a local level component. We placed an inverse-Gamma
prior on its diffusion variance with a prior estimate of $s/\nu= 0.1
\sigma_y$ and a prior sample size $\nu= 32$. The second structural
block was a dynamic regression component. We placed a Gamma prior with
prior expectation $0.1 \sigma_y$ on the diffusion variance of both
regression coefficients. By construction, the outcome variable did not
exhibit any local trends or seasonality other than the variation
conveyed through the covariates. This obviated the need to include an
explicit local linear trend or seasonality component in the model.

In a first analysis, we considered the empirical proportion of
simulations in which a causal effect had been detected. When taking into
account only those simulations where the true effect size was greater than
zero, these empirical proportions provide estimates of the sensitivity
of the model w.r.t. the process by which the data were generated.
Conversely, those simulations where the campaign had  no effect
yield an estimate of the model's specificity. In this way, we
obtained the power curve shown in
Figure~\ref{fig:apps:synthetic:inference}(b). The curve shows that, in data
such as these, a~market perturbation leading to a lift no larger than
1\% is missed in about 90\% of cases. By contrast, a perturbation that
lifts market activity by 25\% is correctly detected as such in most cases.

In a second analysis, we assessed the coverage properties of the
posterior probability intervals obtained through our model.
It is desirable to use a diffuse prior on the local level
component such that central 95\% intervals contain ground truth in
about 95\% of the simulations. This coverage frequency should hold
regardless of the length of the
campaign period. In other words, a longer campaign should lead to
posterior intervals that are appropriately widened to retain the same
coverage probability as the narrower intervals obtained for shorter
campaigns. This was approximately the case throughout the
simulated campaign [Figure~\ref{fig:apps:synthetic:inference}(c)].

\subsubsection*{Estimation accuracy}

\begin{figure}

\includegraphics{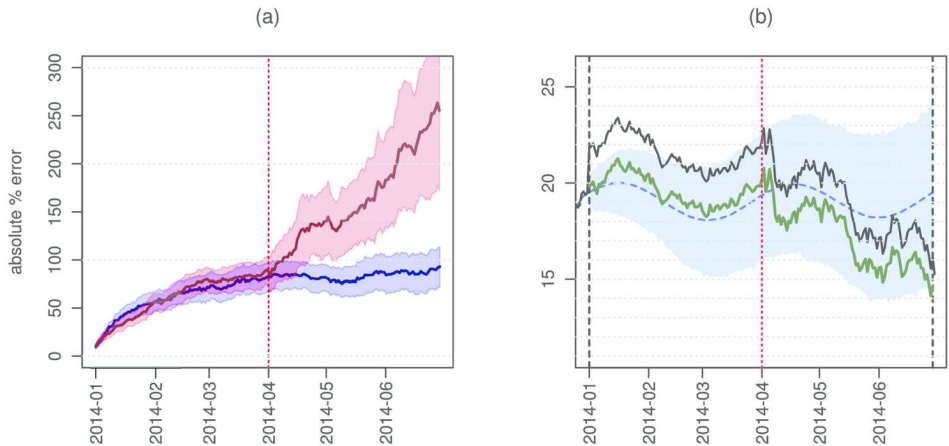}

\caption{Estimation accuracy. \textup{(a)}~Time series of absolute
percentage discrepancy between inferred effect and true effect.
The plot shows the rate (mean${}\pm{}$2 s.e.m.) at which predictions
become less accurate
as the length of the counterfactual forecasting period increases
(blue). The
well-behaved decrease in estimation accuracy breaks down when the
data are subject to a sudden structural change (red), as simulated
for 1~April 2014. \textup{(b)}~Illustration of a structural break. The
plot shows one example of the time series underlying the red curve
in \textup{(a)}. On 1~April 2014, the standard deviation of the generating
random walk of the local level was tripled,
causing the rapid decline in estimation accuracy seen in
the red curve in \textup{(a)}.}
\label{fig:apps:synthetic:acc}
\end{figure}

To study the accuracy of the point estimates supported by our approach, we
repeated the preceding simulations with a fixed effect size of 10\% while
varying the length of the campaign.
When given a quadratic loss function, the loss-minimizing point estimate
is the posterior expectation of the predictive density over counterfactuals.
Thus, for each generated data set $i$, we computed the expected causal effect
for each time point,
%
\begin{equation}
\qquad\hat{\phi}_{i,t} := \langle\phi_t \vert
y_1, \ldots, y_m, x_1, \ldots,
x_m \rangle
\qquad \forall t = n+1, \ldots, m; i = 1, \ldots, 256.\hspace*{-12pt}
\end{equation}
To quantify the discrepancy between estimated and true
impact, we calculated the absolute percentage estimation error,
%
\begin{equation}
a_{i,t} := \frac{\llvert \hat{\phi}_{i,t} - \phi_t \rrvert }{
\phi_t}.
\end{equation}
This yielded an empirical distribution
of absolute percentage estimation errors [Figure~\ref{fig:apps:synthetic:acc}(a), blue], showing
that impact estimates become less and less accurate as the forecasting
period increases. This is because, under the local linear trend model
in (\ref{eq:local-linear-trend}), the true counterfactual activity
becomes more and more likely to deviate from its expected trajectory.
%

%

It is worth emphasising that all preceding results are based on
the assumption that the model structure remains intact throughout the
modelling period. In other words, even though the model is built
around the idea of multiple (nonstationary) components (i.e., a
time-varying local trend and, potentially, time-varying regression
coefficients), this structure itself remains unchanged. If the model
structure does change, estimation accuracy may suffer.

We studied the impact of a changing model structure in a second
simulation in which we repeated the procedure above in such a way that
90~days after the beginning of the campaign the standard deviation of
the random walk governing the evolution of the regression coefficient
was tripled (now 0.03 instead of 0.01). As a result, the observed
data began to diverge much more quickly than before. Accordingly,
estimations became considerably less reliable
[Figure~\ref{fig:apps:synthetic:acc}(a), red]. An example of the underlying
data is shown in Figure~\ref{fig:apps:synthetic:acc}(b).

The preceding simulations highlight the importance of a model that is
sufficiently flexible to account for phenomena
typically encountered in seasonal empirical data. This rules out
entirely static models in particular (such as multiple linear regression).

\section{Application to empirical data}
\label{sec:apps:empirical}

To illustrate the practical utility of our approach, we analysed an
advertising campaign run by one of Google's advertisers in the United
States. In particular, we inferred the campaign's causal effect on the
number of times a user was directed to the advertiser's website from
the Google search results page. We provide a brief overview of the
underlying data below [see \citet{vavermeasuring2011} for additional
details].


The campaign analysed here was based on product-related ads
to be displayed alongside Google's search results for specific
keywords. Ads went live for a period of 6~consecutive weeks and were
geo-targeted to a randomised set of 95 out of 190 designated market
areas (DMAs).
%
The most salient observable characteristic of DMAs is offline sales. To
produce balance in this characteristic, DMAs were first rank-ordered by
sales volume. Pairs of regions were then randomly assigned to treatment/control.
DMAs provide units that can be easily supplied with distinct offerings,
although this fine-grained split was not a requirement for the model.
In fact, we carried out the analysis as if only one treatment region
had been available (formed by summing all treated DMAs). This allowed
us to evaluate whether our approach would yield the same results as
more conventional treatment-control comparisons would have done.

The outcome variable analysed here was search-related visits to the
advertiser's website, consisting of organic clicks (i.e., clicks on a
search result) and paid clicks (i.e., clicks on an ad next to the
search results, for which the advertiser was charged). Since paid
clicks were zero before the campaign, one might wonder why we could not
simply count the number of paid clicks after the campaign had started.
The reason is that paid clicks tend to cannibalise some organic clicks.
Since we were interested in the net effect, we worked with the total
number of clicks.

The first building block of the model used for the analyses in this
section was a local level component. For the inverse-Gamma prior on its
diffusion variance we used a prior estimate of $s/\nu= 0.1 \sigma_y$
and a prior sample size $\nu= 32$. The second structural block was a
static regression component. We used a spike-and-slab prior\vspace*{1pt} with an
expected model size of $M = 3$, an expected explained variance of $R^2
= 0.8$ and 50 prior $df$. We deliberately kept the model as simple as
this. Since the covariates came from a randomised experiment, we
expected them to already account for any additional local linear trends
and seasonal variation in the response variable. If one suspects that a
more complex model might be more appropriate, one could optimise model
design through Bayesian model selection. Here, we focus instead on
comparing different sets of covariates, which is critical in
counterfactual analyses regardless of the particular model structure
used. Model estimation was carried out using 10{,}000 MCMC samples.

%
%
%
\subsection*{Analysis 1: Effect on the treated, using a randomised control}

We began by applying the above model to infer the causal effect of the
campaign on the time series of clicks in the treated regions.
Given that a set of unaffected
regions was available in this analysis, the best possible set of
controls was
given by the untreated DMAs themselves (see below for a comparison
with a purely observational alternative).

%
\begin{figure}

\includegraphics{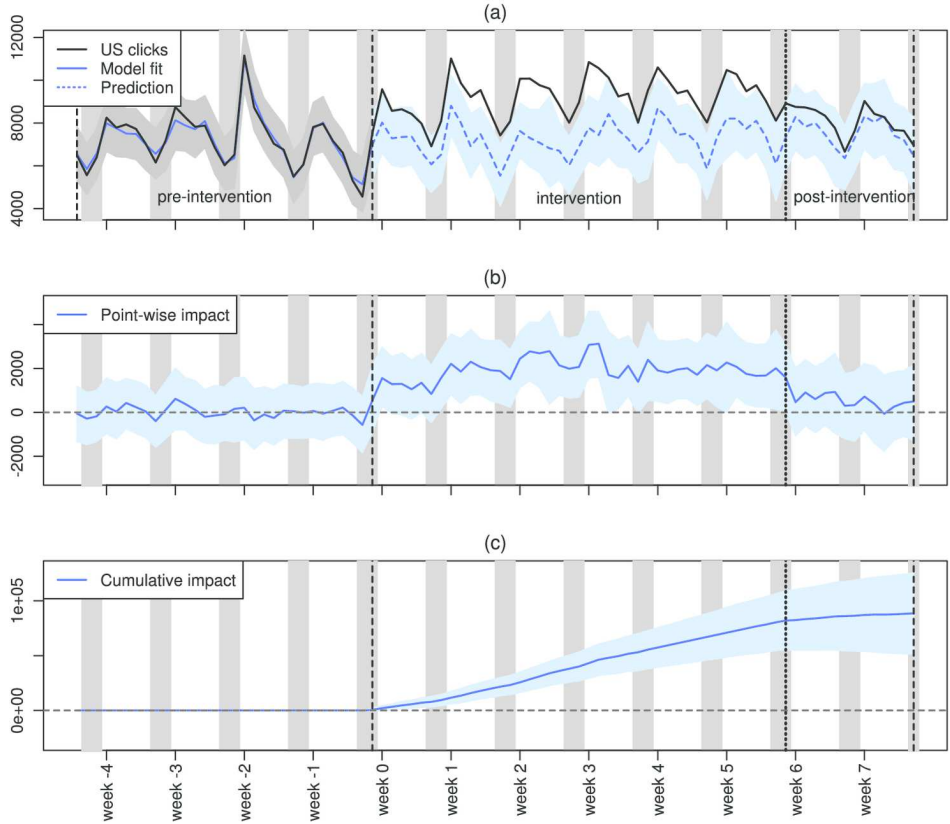}

\caption{Causal effect of online advertising on clicks in treated regions.
\textup{(a)}~Time series of search-related visits to the
advertiser's website (including both organic and paid clicks).
\textup{(b)}~Pointwise (daily) incremental impact
of the campaign on clicks. Shaded vertical bars
indicate weekends. \textup{(c)}~Cumulative impact of the campaign on
clicks.}
\label{fig:apps:empirical:forecast:original}
\end{figure}

As shown in Figure~\ref{fig:apps:empirical:forecast:original}(a), the model
provided an excellent fit on the pre-campaign trajectory of clicks
(including a spike in ``week $-2$'' and a dip at the end of ``week $-1$'').
Following the onset of the campaign, observations quickly began to
diverge from counterfactual predictions: the actual number of clicks
was consistently higher than what would have been expected in the
absence of the campaign. The curves did not reconvene until
one week after the end of the campaign. Subtracting observed from
predicted data, as we did in
Figure~\ref{fig:apps:empirical:forecast:original}(b), resulted in a
posterior estimate of the incremental lift caused by the campaign. It
peaked after about three weeks into the campaign, and faded away after
about one week after the end of the campaign. Thus, as shown in
Figure~\ref{fig:apps:empirical:forecast:original}(c), the campaign led
to a
sustained cumulative increase in total clicks (as opposed to a mere
shift of future clicks into the present or a pure cannibalization of
organic clicks by paid clicks). Specifically, the overall effect
amounted to 88{,}400 additional
clicks in the targeted regions (posterior expectation; rounded to three
significant digits), that is, an
increase of $22\%$, with a central 95\% credible interval of $[13\%,
30\%]$.

To validate this estimate, we returned to the original experimental data,
on which a conventional treatment-control comparison had been carried out
using a two-stage linear model [\citet{vavermeasuring2011}].
This analysis had led to an estimated lift of 84{,}700 clicks, with a
95\% confidence interval for the relative
expected lift of $[19\%,  22\%]$. Thus, with a deviation of less than
5\%, the counterfactual approach had led to
almost precisely the same estimate as the randomised evaluation,
except for its wider intervals. The latter is expected, given that our intervals
represent prediction intervals, not confidence intervals. Moreover, in addition
to an interval for the sum over all time points, our approach
yields a full time series of pointwise intervals, which allows analysts to
examine the characteristics of the temporal evolution of attributable impact.

The posterior predictive intervals in Figure~\ref{fig:apps:empirical:forecast:original}(b) widen more slowly than in the
illustrative example in Figure~\ref{fig:intro:illustration}. This is
because the large number of controls available in this data set offers
a much higher pre-campaign predictive strength than in the simulated
data in Figure~\ref{fig:intro:illustration}. This is not unexpected, given
that controls came from a randomised experiment, and we will see that
this also holds for a subsequent analysis (see below) that is based on
yet another data source for predictors. A consequence of this is that
there is little variation left to be captured by the random-walk
component of the model. A reassuring finding is that the estimated
counterfactual time series in Figure~\ref{fig:apps:empirical:forecast:original}(a) eventually almost exactly
rejoins the observed series, only a few days after the end of the intervention.

\subsection*{Analysis 2: Effect on the treated, using observational controls}

An important characteristic of counterfactual-forecasting approaches is that
they do not require a setting in which a set of controls, selected at
random, was exempt from the campaign. We therefore repeated the preceding
analysis in the following way:
%
we discarded the data from all control
regions and, instead, used searches for keywords related to the
advertiser's industry, grouped into a handful of verticals, as covariates.
In the absence of a dedicated set
of control regions, such industry-related time series can be very powerful
controls, as they capture not only seasonal variations but also market-specific
trends and events (though not necessarily advertiser-specific trends).
%
A major strength of the controls chosen here is that time series on web
searches are publicly available through Google Trends
(\url{http://www.google.com/trends/}). This makes the approach applicable to
virtually any kind of intervention. At the same time, the industry as a
whole is unlikely to be moved by a single actor's activities. This
precludes a positive bias in estimating the effect of the campaign that
would arise if a covariate was negatively affected by the campaign.

\begin{figure}

\includegraphics{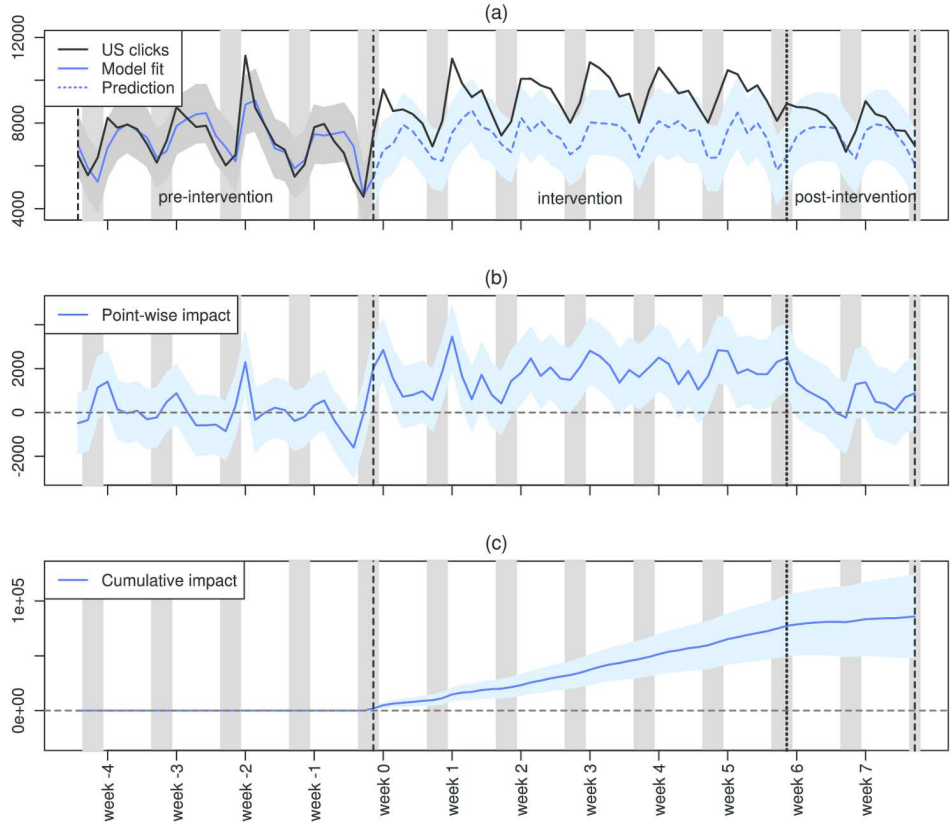}

\caption{Causal effect of online advertising on clicks, using only
\emph{searches} for keywords related to the advertiser's industry as
controls, discarding the original control regions
as would be the case in studies where a randomised
experiment was not carried out. \textup{(a)}~Time series of clicks on
to the advertiser's website. \textup{(b)}~Pointwise
(daily) incremental impact of the campaign on clicks.
\textup{(c)}~Cumulative impact of the campaign on clicks. The plots show
that this analysis, which was based on observational covariates
only, provided almost exactly the same inferences as the first
analysis (Figure~\protect\ref{fig:apps:empirical:forecast:original}) that had
been based on a randomised design.}
\label{fig:apps:empirical:forecast:observational}
\end{figure}

As shown in Figure~\ref{fig:apps:empirical:forecast:observational}, we
found a cumulative lift of 85{,}900 clicks (posterior expectation), or
$21\%$, with a $[12\%,  30\%]$ interval.
In other words, the analysis replicated almost perfectly the original
analysis that had
access to a randomised set of controls. One feature in the
response variable which this second analysis failed to account for was
a spike in clicks in the second week before the campaign onset; this spike
appeared both in treated and untreated regions and appears to be
specific to
this advertiser. In addition, the series of point-wise impact
[Figure~\ref{fig:apps:empirical:forecast:observational}(b)] is slightly
more volatile
than in the original analysis (Figure~\ref{fig:apps:empirical:forecast:original}). On the other hand, the overall
point estimate of 85{,}900, in this case, was even closer to the
randomised-design baseline (84{,}700; deviation ca.~1\%) than in our
first analysis (88{,}400; deviation ca.~4\%). In summary, the
counterfactual approach effectively obviated the need for the original
randomised experiment. Using
purely observational variables led to the same substantive conclusions.

\subsection*{Analysis 3: Absence of an effect on the controls}

To go one step further still, we analysed clicks in those regions
that had been exempt from the advertising campaign. If the effect of
the campaign was truly specific to treated regions, there should be no
effect in the controls. To test this, we inferred the causal effect of
the campaign on \emph{unaffected}
regions, which should \emph{not} lead to a significant finding.
In analogy with our second analysis, we discarded clicks in the treated
regions and used searches for keywords related to the advertiser's
industry as controls.

%
\begin{figure}

\includegraphics{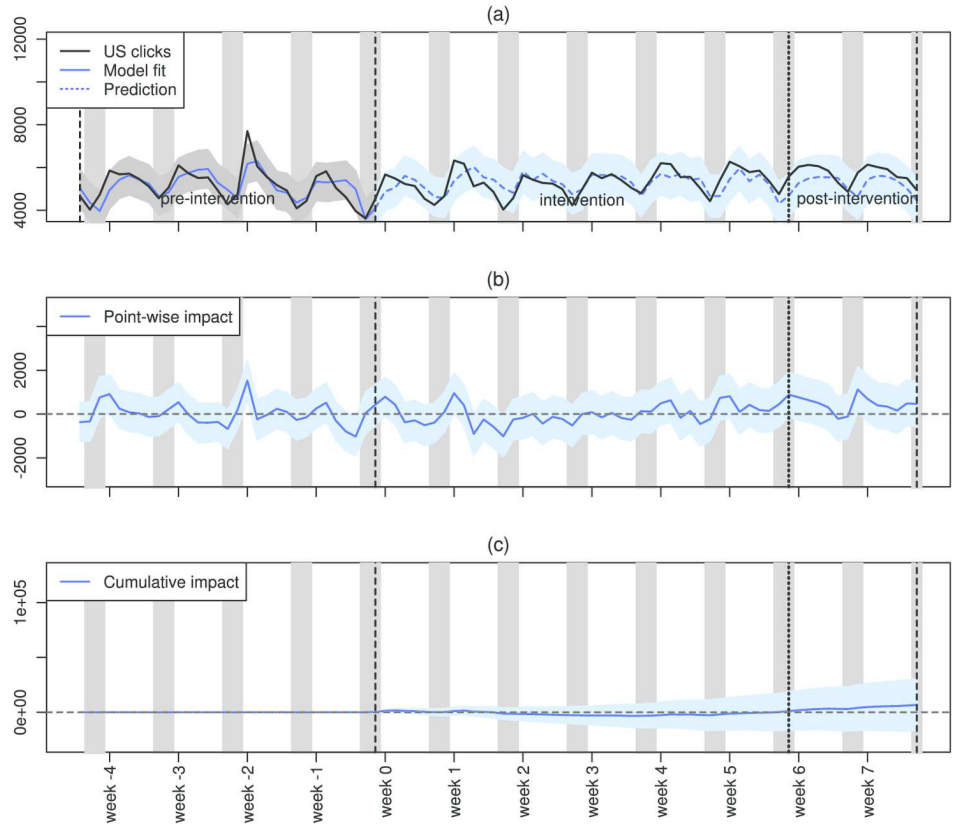}

\caption{Causal effect of online advertising on clicks in
\emph{nontreated} regions, which should \emph{not} show an
effect. Searches for keywords related to the advertiser's industry are
used as controls. Plots show
inferences in analogy with
Figure~\protect\ref{fig:apps:empirical:forecast:original}. \textup{(a)}~Time series of
clicks to the advertiser's website.
\textup{(b)}~Pointwise (daily) incremental impact of the campaign on
clicks.
\textup{(c)}~Cumulative impact of the campaign on clicks.}
\label{fig:apps:empirical:forecast:null}
\end{figure}

As summarised in Figure~\ref{fig:apps:empirical:forecast:null}, no
significant effect was found in unaffected regions, as expected.
Specifically, we obtained an overall nonsignificant lift of $2\%$ in
clicks with a central 95\% credible interval of $[-6\%,  10\%]$.

In summary, the empirical data considered in this section showed:
(i)~a clear effect of advertising on treated regions when using
randomised control regions to form the regression component, replicating
previous treatment-control comparisons
(Figure~\ref{fig:apps:empirical:forecast:original});
(ii)~notably, an equivalent finding when discarding control regions and
instead using observational searches for keywords related to the
advertiser's industry as covariates
(Figure~\ref{fig:apps:empirical:forecast:observational});
(iii)~reassuringly, the absence of an effect of advertising on regions that
were not targeted (Figure~\ref{fig:apps:empirical:forecast:null}).

\section{Discussion}
\label{sec:disc}

The increasing interest in evaluating the incremental impact of market
interventions has been reflected by a growing literature on
applied causal inference. With the present paper we
are hoping to contribute to this literature by proposing a Bayesian
state-space model for obtaining a counterfactual prediction of market
activity. We discuss the main features of this model below.

In contrast to most previous schemes, the approach described here is fully
Bayesian, with regularizing or empirical priors for all
hyperparameters. Posterior inference gives rise to complete-data
(smoothing) predictions that are only conditioned on past data in the
treatment market and both past and present data in the control
markets. Thus, our model embraces a dynamic evolution of states and,
optionally, coefficients (departing from classical linear regression
models with a fixed number of static regressors), and enables us to
flexibly summarise posterior inferences.

Because closed-form posteriors for our model do not exist, we suggest
a stochastic approximation to inference using MCMC. One convenient
consequence of this is that we can reuse the samples from the
posterior to obtain credible intervals for all summary statistics of
interest. Such statistics include, for example, the average absolute
and relative effect caused by the intervention as well as its
cumulative effect.

Posterior inference was implemented in C$++$ and {R} and, for all
empirical data sets presented in Section~\ref{sec:apps:empirical},
took less
than 30~seconds on a standard Linux machine. If the computational
burden of sampling-based inference ever became prohibitive, one
option would be to replace it by a variational Bayesian approximation
[see, e.g., \citet{mathysbayesian2011,brodersenvariational2013}].

Another way of using the proposed model is for power analyses. In
particular, given past time series of market activity, we can define a
point in the past to represent a hypothetical
intervention and apply the model in the usual fashion. As a result, we
obtain a measure of uncertainty about the response in the treated
market after the beginning of the hypothetical intervention. This
provides an estimate of what incremental effect would have
been required to be outside of the 95\% central interval of what would
have happened in the absence of treatment.

The model presented here subsumes several simpler models which, in consequence,
lack important characteristics, but which may
serve as alternatives should the full model appear too complex for the
data at hand.
%
One example is classical multiple linear regression. In principle,
classical regression models go beyond difference-in-differences
schemes in that they account for the full counterfactual
trajectory.
%
However, they are not suited for predicting stochastic
processes beyond a few steps. This is because ordinary least-squares
estimators disregard serial autocorrelation; the static model
structure does not allow for temporal variation in the
coefficients; and predictions ignore our posterior uncertainty
about the parameters.
Put differently: classical multiple linear regression is a special
case of the state-space model described here in which (i)~the Gaussian
random walk of the local level has zero variance; (ii)~there is no
local linear trend; (iii)~regression coefficients are static rather
than time-varying; (iv)~ordinary least squares estimators are used
which disregard posterior uncertainty about the parameters and may easily
overfit the data.

Another special case of the counterfactual approach discussed in this
paper is given by synthetic control estimators that are restricted to
the class of convex combinations of predictor variables and do not
include time-series effects such as trends and seasonality [\citet{abadiesynthetic2010,abadiesemiparametric2005}]. Relaxing this
restriction means we can utilise predictors regardless of their scale,
even if they are negatively correlated with the outcome series of the
treated unit.

Other special cases
include autoregressive (AR) and moving-average (MA) models. These
models define autocorrelation among observations rather than latent
states, thus precluding the ability to distinguish between state noise
and observation noise [\citet{atamanbuilding2008,leeflangcreating2009}].

In the scenarios we consider, advertising is a planned
perturbation of the market. This generally makes it easier to obtain
plausible causal inferences than in genuinely \emph{observational}
studies in which the experimenter had no control about treatment
[see discussions in \citet{antonakismaking2010,berndtpractice1991,bradymodels2002,camillonew2010,hitchcockall2004,kleinbergreview2011,%
lewisdoes2011,lewishere2011},
Robinson, McNulty and Krasno
(\citeyear{robinsonobserving2009}),
\citet{vavermeasuring2011,winshipestimation1999}]. The principal problem in observational
studies is endogeneity: the possibility that the observed outcome
might not be the result of the treatment but of other omitted,
endogenous variables. 
In principle, propensity scores can be used to correct for the selection
bias that arises when the treatment effect is correlated with the
likelihood of being treated
[\citet{rubinestimating2006,chanevaluating2010}]. However, the
propensity-score approach requires that exposure can be measured at
the individual level, and it, too, does not guarantee valid inferences,
for example, in the presence of a specific type of selection bias recently
termed ``activity bias'' [\citet{lewishere2011}].
%
Counterfactual modelling approaches avoid these issues when it can
be assumed that the treatment market was chosen at random.

Overall, we expect inferences on the causal impact of designed market
interventions to play an increasingly prominent role in providing quantitative
accounts of return on investment
[\citet{danaherdetermining1996,seggiemeasurement2007,leeflangcreating2009,stewartmarketing2009}].
This is because marketing resources, specifically, can only be allocated
to whichever campaign elements jointly provide the greatest return on
ad spend
(ROAS) if we understand the causal effects of spend on sales, product
adoption or user engagement.
At the same time, our approach could be used for many other
applications involving causal inference. Examples include problems
found in economics, epidemiology, biology or the political and social
sciences. With the release of the {\tt CausalImpact} R package we hope
to provide a simple framework serving all of these areas.
Structural time-series models are being used in an increasing number of
applications at Google, and we anticipate that they will prove equally
useful in many analysis efforts elsewhere.

\section*{Acknowledgment}

The authors wish to thank Jon Vaver for sharing the empirical data
analysed in this paper.




\printaddresses
\end{document}